\documentclass[a4paper]{article}
\usepackage{amsmath,amssymb}
\usepackage{color}
\usepackage{epsfig,graphicx}
\usepackage[T1]{fontenc}
\usepackage{lmodern}
\usepackage[english]{babel}
\usepackage{amsthm}
\usepackage{array}
\usepackage{enumerate}
\usepackage{fancyhdr}
\usepackage{graphicx}
\usepackage[usenames,dvipsnames]{xcolor}
\usepackage{float}
\usepackage{epstopdf}
\usepackage{booktabs}
\usepackage{tikz}
\usepackage{fancyref}
\usepackage{caption}
\usepackage{mathrsfs}
\usepackage{hyperref}
\usepackage{url}

\newcommand{\imatge}[5]{
\begin{figure}
[H]
\centering
\includegraphics[scale=#2, angle=#3]{#1}
\caption{\small{#4}}
\label{#5}
\end{figure}
}

\theoremstyle{definition}
\usepackage{amsthm}
\newtheorem{thm}{Theorem}[subsection]

\newtheorem{exemple}[thm]{Example}
\def\GG{\mathcal G}
\def\LL{\mathcal L}

\begin{document}

\title{Gauss-Bonnet modified gravity models with bouncing behavior}
\author{Anna Escofet\footnote{E-mail address:
anna.escofetp@gmail.com} \, {\small and} \, Emilio Elizalde\footnote{E-mail address:
elizalde@ieec.uab.es} \\
\small ICE-CSIC and IEEC, UAB Campus, C/ Can Magrans s/n,
 08193 Bellaterra (Barcelona) Spain}

\maketitle

\thispagestyle{empty}

\begin{abstract}
The following issue is addressed: how the addition of a Gauss-Bonnet term (generically coming from most fundamental theories, as string and M theories), to a viable model, can change the specific properties, and even the physical nature, of the corresponding cosmological solutions? Specifically, brand new original dark energy models are obtained in this way with quite interesting properties, which exhibit, in a unified fashion, the three distinguished possible cosmological phases corresponding to phantom matter, quintessence, and ordinary matter, respectively. A model, in which the equation of state parameter, $w$, is a function of time, is seen to lead either to a singularity of the Big Rip kind or to a bouncing solution which evolves into a de Sitter universe with $w=-1$. Moreover, new Gauss-Bonnet modified gravity models with bouncing behavior in the early stages of the universe evolution are obtained and tested for the validity and stability of the corresponding solutions. They allow for a remarkably natural, unified description of a bouncing behavior at early times and accelerated expansion at present.
\end{abstract}

\section{Introduction}

The key issue we address in this work is how the addition of a Gauss-Bonnet term to some known, working models can change the specific properties---and even the physical nature---of the corresponding cosmological solutions. As is well known, the standard Big Bang model provides answers to three main questions: the universe expansion, observed through the redshift of distant galaxies, the cosmic background radiation, and the relative abundances of the light elements. A good model for the universe should also, on its turn, solve three noted problems, namely the flatness, horizon and monopole problems. The good news we want to make aware of with the present work are that, in order to deal with all these issues, one can employ modified gravity models based on string theory inspired loop quantum cosmology and the ekpyrotic scenario. In our paper we will start from some of these models and a number of new ones will be constructed---by using the already mentioned procedure---which will have well behaved, stable solutions.

We will start by revisiting a model \cite{Nojiri2} that explains accelerated expansion and which (this is our actual contribution here) could lead to an oscillating solution when using Gauss-Bonnet modified gravity. More specifically, we will consider a model with a scalar field, $\phi$, coupled to gravity and, as a correction to the standard action and inspired by string and M theory, we will add to it a term proportional to the Gauss-Bonnet invariant (a mathematically unique, well-grounded term). When the potential is zero and the usual parameter  $\gamma$ is canonical, we will show that the new model exhibits three cosmological phases corresponding to phantom matter, quintessence, and ordinary matter, respectively. In the case when the potential is positive, it is also possible to find solutions with a constant equation of state (EoS) parameter, $w$. Also, a new model in which $w$ is not constant will be considered. This model is such that when the curvature is small, we get $w<-1$, while when it is large, we get $w>-1$, so that within this model we can either have a Big Rip singularity or a bouncing solution tending to a de Sitter universe with $w=-1$, thus avoiding the singularity. It should also mentioned that the possibility of avoiding the curvature
singularity thanks to Gauss-Bonnet corrections was also considered, in the context of string cosmology, in previous papers such as \cite{str3,str3b,str3c}.

A competitive alternative to the standard inflationary description of the early acceleration stage is provided by bouncing cosmologies. Another key issue in our paper, in this respect, will be the investigation of some Gauss-Bonnet modified gravity models with a bouncing behavior (for some relevant works on bouncing cosmologies, see \cite{Novello,Li,Cai,Cai2,Haro2,Qiu,Qiu2,Lehners,Cai3,Odintsov2,Khoury2}).
Our goal in this case will be to reconstruct, using Gauss-Bonnet modified gravity models, the superbounce and loop quantum ekpyrotic cosmologies. That is, given the scale factor, to find the model that realizes these cosmologies and to study the stability of the solutions found. Our final results will have the form of two new models: a superbounce model which is stable and a model for the ekpyrotic scenario which is not. This non-singular bouncing cosmology and its stability will be studied in all detail as well.

The paper is divided into two main parts. In the first one, Sect.~2, making use of the procedure explained above, two new original dark energy models are constructed which exhibit, in a unified way, the three distinguished possible cosmological phases corresponding to phantom matter, quintessence, and ordinary matter, respectively. In addition, a third model is presented, where one can either obtain a singularity of the Big Rip kind or a bouncing solution tending to a de Sitter universe with $w=-1$. The fact that such different cosmological behaviors can be unified, as coming from a single cosmological model needs to be properly stressed. In the second part of the paper, Sect.~3, we construct Gauss-Bonnet modified gravity models with a bouncing behavior in the early stages of the universe evolution, and test the stability of the corresponding solutions. A superbounce model is presented which is stable, as well as a model for the ekpyrotic scenario which is not. Finally, Sect.~4 is devoted to Conclusions.

\section[Dark energy in Gauss-Bonnet modified gravity]{Dark energy in Gauss-Bonnet modified gravity}

In this section our aim is to demonstrate how accelerated expansion can be explained via Gauss-Bonnet modified gravity. To start, we consider a model with a scalar field, $\phi$, coupled with gravity to which action we will add a term proportional to the Gauss-Bonnet invariant, as a correction inspired by string theory and M theory \cite{str3,str3b,str3c,Cognola,Cognolab,Elizalde,Elizaldeb,Elizaldec,Elizalded,Nojiri}, and we will investigate the properties of these new model.

Present accelerated expansion and flatness are usually understood by accepting the existence of dark energy. Accurate astronomical observations indicate that the energy density of the universe must be very close to its critical density. However, the total matter density in the universe (including barionic and dark matter) represents only approximately 30\% of the critical density, thus it must exist an additional and unknown energy form to account for the remaining 70\%.
The, in principle, simplest possibility suggested for dark energy is the cosmological constant, a constant energy density filling the space homogeneously and which was already considered by Einstein as a purely mathematical possibility to render the universe stationary, by countering gravity collapse. At present we can more reasonably consider it as an energy coming from the unavoidable fluctuations of the quantum vacuum (i.e. as the price to pay for `having space'), which do show up in any quantum theory as a result of the uncertainty principle. On the other hand, scalar fields, as dynamical quantities whose energy density can vary with time and space, have also been postulated. Nevertheless, some models imply scalar fields varying so subtly that distinguishing them from the cosmological constant, or even among themselves, is almost impossible with the astronomical data we have at hand.

Dark energy, according to the most accurate interpretation of observations, is characterized by having negative pressure and an EoS parameter, $w$, oscillating apparently very close to $-1$ (and maybe a bit below this value). For large time scales the existence of dark energy seems clear; however, building a quantum cosmological model which uses it consistently, at any scale, is a big and difficult challenge. In the new models here considered, Gauss-Bonnet modified gravity will play a key role since it will endow the solutions with additional and very interesting properties.

\subsection{Model description}
Let us consider the Einstein-Hilbert action
\begin{eqnarray}
 S=\int d^4x\sqrt{-g}\left(\frac{R}{2\kappa^2}-\frac{\gamma}{2}\partial_\mu\phi\partial^\mu\phi-V(\phi),
 +f(\phi)\GG+\LL_{matter}\right),
\end{eqnarray}
where $R$ is the scalar curvature, $\GG=R^2-4R_{\mu\nu}R^{\mu\nu}+
R_{\mu\nu\rho\sigma}R^{\mu\nu\rho\sigma}$ is the Gauss-Bonnet invariant and $\gamma=\pm 1$.

Varying this action with respect to $\phi$,  we obtain
\begin{eqnarray}
 0=\gamma\nabla^2\phi-V'(\phi)+f'(\phi)\GG,
\end{eqnarray}
and varying with respect to the metric,
\begin{equation}
\begin{split}
 0&=\frac{1}{\kappa^2}\left(-R^{\mu\nu}+\frac{1}{2}g^{\mu\nu}R\right)+\gamma
 \left(\frac{1}{2}\partial^\mu\phi\partial^\nu\phi-\frac{1}{4}g^{\mu\nu}
 \partial_\rho\phi\partial^\rho\phi\right)+\frac{1}{2}g^{\mu\nu}(-V(\phi)+f(\phi)\GG)\\
 &-2f(\phi)RR^{\mu\nu}+2\nabla^\mu\nabla\nu(f(\phi)R)-2g^{\mu\nu}\nabla^2(f(\phi)R)\\
 &+8f(\phi)R^\mu_\rho R^{\nu\rho}-4\nabla_\rho\nabla^\mu(f(\phi)R^{\mu\rho})-4\nabla_\rho\nabla^\nu(f(\phi)R^{\mu\rho})\\
 &+4\nabla^2(f(\phi)R^{\mu\nu})+4g^{\mu\nu}\nabla_\rho\nabla_\sigma(f(\phi)R^{\rho\sigma})
 -2f(\phi)R^{\mu\rho\sigma\tau}R^\nu_{\rho\sigma\tau}\\
 &+4\nabla_\rho\nabla_\sigma(f(\phi)R^{\mu\rho\sigma\tau}).
\end{split}
\end{equation}

Let us consider the FLRW metric, and suppose that $\phi$ only depends on time, then
\begin{equation}\label{eq1}
0=-\frac{3}{\kappa^2}H^2+\frac{\gamma}{2}\dot{\phi}^2+V(\phi)-24\dot{\phi}f'(\phi)H^3
\end{equation}
and
\begin{equation}\label{eq2}
 0=-\gamma(\ddot{\phi}+3H\dot{\phi})-V'(\phi)+24f'(\phi)(\dot{H}H^2+H^4).
\end{equation}
\subsection{Particular case: Accelerated universe}
Take now the particular case
\begin{equation}\label{vif}
 V(\phi)=V_0e^{-2\frac{\phi}{\phi_0}},~~~~~~~~~~ f(\phi)=f_0e^{2\frac{\phi}{\phi_0}},
\end{equation}
and suppose that the scale factor is
\begin{equation}
a(t)=a_0t^{h_0}.
\end{equation}
\begin{itemize}
\item If $h_0>0$, then the scale factor corresponds to an expanding universe.
\item If $h_0<0$, then the scale factor corresponds to a shrinking universe. Changing the direction of time, we have an expanding universe with $a=a_0(-t)^{h_0}$. Changing the origin of time $t\to t_s-t$, then $t$ is positive as long as $t<t_s$.
\end{itemize}
Consider the following behavior for the Hubble rate,
\begin{eqnarray}\label{hs}
\begin{cases}
H=\frac{h_0}{t},~~\phi=\phi_0\ln\left(\frac{t}{t_1}\right), ~~~~~~~~~~~~\text{ if }h_0>0,\\
H=-\frac{h_0}{t_s-t},~~\phi=\phi_0\ln\left(\frac{t_s-t}{t_1}\right), ~~~~~\text{ if }h_0<0,\\
\end{cases}
\end{eqnarray}
and, from Eqs.~(\ref{eq1}) and (\ref{eq2}), we obtain, in both cases,
\begin{equation}\label{eq11}
 V_0t_1^2=-\frac{1}{\kappa^2(1+h_0)}\left(3h_0^2(1-h_0)+\frac{\gamma\phi_0^2\kappa^2(1-5h_0)}{2}\right),
 \end{equation}
 \begin{equation}\label{eq22}
\frac{48f_0h_0^2}{t_1^2}=-\frac{6}{\kappa^2(1+h_0)}\left(h_0-\frac{\gamma\phi_0^2\kappa^2}{2}\right).
 \end{equation}
Note that if $-1<h_0<0$ and $\gamma=1$, then it has to be $f_0<0$. Note also that
suppressing the Gauss-Bonnet term effect, setting $f(\phi)=0$, then
\begin{equation}\label{XXX}
 h_0=\frac{\gamma\phi_0^2\kappa^2}{2}.
\end{equation}
The equation of state is $w=-1+\frac{2}{3h_0}$, thus
\begin{itemize}
 \item $h_0<0$ $\Longrightarrow$ $w<-1$.
 \item $h_0>0$ $\Longrightarrow$ $w>-1$.
\end{itemize}
From Eqs.~(\ref{eq11}) and (\ref{eq22}) it follows that we can choose appropriate parameters in order to get $h_0<0$, and then we obtain phantom matter.

\begin{exemple}
 Consider the case $V(\phi)=0$. In the case (\ref{vif}), $V(\phi)=0$ if and only if $V_0=0$, and from Eq.~(\ref{eq22}), we know that $V_0=0$ if and only if
 \begin{equation}
  \phi_0^2=-\frac{6h_0^2(1-h_0)}{\gamma(1-5h_0)\kappa^2}.
 \end{equation}
When $\gamma=1$, for $\phi_0$ to be real it must be
$\frac{1}{5}<h_0<1$. When $\gamma=-1$, for $\phi_0$ to be real it must be $h_0<\frac{1}{5}$ or $h_0>1$.


\begin{itemize}
 \item[*] When $\gamma=1$, there is a solution with $\frac{1}{5}<h_0<1$.
 \item[*] When $\gamma=-1$, there are three solutions:
 \subitem when $h_0<0$, the solution describes a universe with phantom matter.
 \subitem when $h_0>1$, the solution describes a universe with quintessence.
 \subitem when $0<h_0<\frac{1}{5}$, the solution describes ordinary matter.
\end{itemize}
In the case  $\gamma=1$ and $V(\phi)=0$, we cannot obtain a universe with phantom matter with $h_0<0$, or $w<-1$.
 \end{exemple}

\begin{exemple}
The particular case of (\ref{vif}): $\phi=\varphi_0$, $H=H_0$ also solves (\ref{eq1}) and (\ref{eq2}) and a deSitter space. With (\ref{eq1}), (\ref{eq2}) and (\ref{vif}), we find the Hubble parameter to be $H_0^2=-\frac{e^{\frac{2\varphi_0}{\phi_0}}}{8f_0\kappa^2}$, where $f_0<0$ is needed for the solutions to exist, $\varphi$ is arbitrary, and then we have the Hubble constant determined from an initial condition.
\end{exemple}
\subsection{Asymptotic cosmology}

Let us now consider the more general situation,
\begin{equation}\label{vif2}
V(\phi)=V_0e^{-2\frac{\phi}{\phi_0}},~~~~~~~~~~
 f(\phi)=f_0e^{2\frac{\phi}{\alpha\phi_0}}.
\end{equation}
We want to study the asymptotic behavior of the solutions. \medskip

\noindent\textbf{Case $\alpha>1$.}

When the curvature is small, that is $t\to\infty$ or $t_s-t\to\infty$ in (\ref{hs}), the Gauss-Bonnet term can be neglected, and the solution is $h_0=\frac{\gamma\phi_0^2\kappa^2}{2}$. In this case  $h_0<0$ if and only if $\gamma=-1$, and then a universe with $w<-1$ can only happen in the case $\gamma=-1$, in this model.

When the curvature is large, that is $t\to0$ or $t_s-t\to0$ in (\ref{hs}), the classic potential term can be neglected.
Suppose
\begin{eqnarray}\label{by}
\begin{cases}
H=\frac{h_0}{t},~~\phi=\alpha\phi_0\ln\left(\frac{t}{t_1}\right), ~~~~~~~~~~~~\text{ if }h_0>0,\\
H=-\frac{h_0}{t_s-t},~~\phi=\alpha\phi_0\ln\left(\frac{t_s-t}{t_1}\right), ~~~~~\text{ if }h_0<0,\\
\end{cases}
\end{eqnarray}
 then Eqs.~(\ref{eq1}) and (\ref{eq2}) turn into
 \begin{equation}
  0=-\frac{3h_0^2}{\kappa^2}+\frac{\gamma\alpha^2\phi_0^2}{2}-\frac{48f_0h_0^3}{t_1^2},
 \end{equation}
 \begin{equation}
  0=\gamma(1-3h_0)\alpha^2\phi_0^2+\frac{48f_0h_0^3}{t_1^2}(h_0-1),
 \end{equation}
and, with $f_0=0$, we have
\begin{equation}
 \phi_0^2=-\frac{6h_0^2(1-h_0)}{\gamma\alpha^2(1-5h_0)\kappa^2},
\end{equation}
which exhibits the same qualitative behavior as the model (\ref{vif}) when $\gamma=1$.

In the case $\gamma=-1$, when the curvature is small, as in the present universe, the potential term dominates: we have cosmic acceleration with $w<-1$, the curvature increases, and when it is large the Gauss-Bonnet term dominates. Thus, it turns out that  $w>-1$, and the Big Rip singularity is avoided. Then, the curvature decreases again.

Therefore, in this case either a Big Rip singularity can occur or, either, an oscillating solution may appear tending to a de Sitter universe with $w=-1$.

In this model, also when assuming $\phi=\varphi_0$, $H=H_0$, we have a de Sitter solution,
\begin{equation}
H_0^2=\frac{e^{-\frac{2\varphi_0}{\alpha\phi_0}}}{8f_0\kappa^2}, ~~~~~~~ \varphi=\frac{\alpha\phi_0}{2(1-\alpha)}\ln\left(-\frac{8V_0f_0\kappa^2}{3}\right).
\end{equation}

\noindent\textbf{Case $0<\alpha<1$.}

A solution where $h_0>0$ appears with $\gamma=-1$;
however, when $\gamma=1$, there is no solution for an accelerating universe with $w<-1$; in this case $w(t)>-\frac{1}{3}$, for all time $t$.

\section[Bouncing cosmology in Gauss-Bonnet modified gravity]{Bouncing cosmology in Gauss-Bonnet modified gravity}

We start this section with a summary of an alternative scenario and the improvement introduced in this case by the addition of the Gauss-Bonnet term, what is our leading idea, as well as the common procedure, in this paper. General relativity theorizes a universe in which a singularity occurred at some finite time in the past, but it turns out that the theory itself is no longer valid when one closely approaches the singularity that it predicts since, necessarily, quantum effects take over. 
  The enormously high energy density or, equivalently in terms of geometry, extremely large curvature, at the initial stages of the universe formation, sets quantum gravity theory as the best description of the universe at the Big Bang scale \cite{Cognola,Elizalde,Khoury,Ashtekar,Bojowald,Bamba2}; a theory, by the way, whose precise formulation is still lacking. However, quantum corrections to the classical theory of gravity have been formulated in a reasonable way. Summing up, general relativity describes accurately the universe at large scales of time and space, while quantum gravity corrections offer a dynamical description at small scales and also evolution equations for the wave equations describing the first stages of the universe evolution.

It turns out that the introduction of quantum gravity corrections in cosmology result in a very different vision of the universe: according to the classical theory, nothing can prevent matter and energy from collapsing into black holes or the Big Bang singularity when looking backwards in time. In the classical Friedmann equation the potential decreases when the scale factor decreases, so that it diverges when $a=0$. Quantum effects, however, and specifically loop quantum gravity, dictate that the potential starts increasing at some point before arriving at $a=0$.  This implies that, at a certain moment, when the effects of quantum gravity become more important, attractive gravity switches to repulsive gravity, avoiding the collapse.
In a gravitational background a change in the  behavior of matter is observed: in the classical model, the matter fields in an expanding universe are slowed down by a friction term in the Klein-Gordon equation. In a shrinking universe, the fields are excited.
This behavior turns around at quantum scales: the friction in an expanding universe becomes antifriction and this way matter fields are removed from their minimum potential states before the classical behavior can become more important.

Analogously, in a shrinking universe the fields freeze in close proximity of the singularity. Gravitational repulsion does not only exist near the singularity but also at conveniently small scales. The collapse would be avoided in this theory and also the expansion would be stronger. The universe would accelerate its expansion as a direct consequence of gravitational quantum repulsion, eventually matching astronomical observations. These and other mechanisms depict a universe in which, in contrast to the classical theory, the Big Bang would not actually be the beginning of the universe but, instead, expansion and shrinking would alternate in time avoiding the formation of singularities. These cyclic models would however not explain the very genesis of the cosmos, placing it in the infinitely far past.

One of the postulated models in loop quantum cosmology is the ekpyrotic scenario. It uses a view of the universe which is somehow different: it consists of a 5-dimensional space-time in which two 4-dimensional surfaces, or 3-branes, are placed at a finite distance between them, along the fifth spatial dimension. The visible brane would correspond to our observable 4-dimensional universe and the other one, the hidden brane, would remain undetectable. The ekpyrotic universe pretends to give solutions to the main cosmological questions discussed in the Introduction, as the horizon, the monopole, and the flatness problems.

Starting from an apparently cold and stable state in which the two branes would remain stable, a third brane (bulk brane) would secede spontaneously from the hidden brane and move to the visible one. The slow collision between our brane and the bulk brane through the fifth dimension would cause the apparent Big Bang. The universe could have existed during an indefinite amount of time before the collision, but it still makes sense to define this moment as the beginning of cosmic time.  The bulk brane and the visible brane fuse through a transition during which part of the kinetic energy of the bulk brane is converted into a hot thermal bath of radiation and matter on the visible brane. At this point the Big Bang begins, but from a finite temperature state. Note that in this model, in contrast with the classical theory, a singularity with infinite temperature is avoided. Our universe would start expanding from a finite temperature stage.

The name ekpyrotic comes from this view and from the cyclic nature of the system; it evokes the Stoic conception of cosmic evolution of the universe, according to which it is supposed to be consumed by and reborn from flames, repeatedly. In our case, the ignition is caused by the brane collisions along the fifth dimension. Although being flat in average, the bulk brane generates some ripples due to quantum fluctuations, which result in different regions colliding and reheating at slightly different times, thereby impressing a spectrum of density fluctuations on the visible brane. For some other relevant works on bouncing cosmology in Gauss-Bonnet gravity, see \cite{Bamba3,Astashenok,Haro,Bamba4}. We will now describe the model we are going to deal with in the context of Gauss-Bonnet modified gravity.

\subsection{Description of the model}
Consider the Einstein-Hilbert action \cite{Nojiri5}
\begin{eqnarray}
 S=\frac{1}{2\kappa^2}\int d^4x\sqrt{-g}[R+F(\GG)]+S_{matter},
\end{eqnarray}
where $\kappa^2=8\pi G$ is the gravitational constant, $g=\det(g_{\mu\nu})$, $R$ is the scalar curvature,  $\GG=R^2-4R_{\mu\nu}R^{\mu\nu}+R_{\mu\nu\rho\sigma}R^{\mu\nu\rho\sigma}$ the Gauss-Bonnet invariant, and $S_{matter}$ is the matter action.

For an arbitrary metric, the corresponding field equations are
\begin{equation}
\begin{split}
&R_{\mu\nu}-\frac{1}{2}g_{\mu\nu}F(\GG)+\left(2RR_{\mu\nu}-4R_{\mu\rho}R_\mu^\rho+
2R_\mu^{\rho\sigma\tau}R_{\nu\rho\sigma\tau}-4g^{\alpha\rho}g^{\beta\sigma}R_{\mu\alpha\nu\beta}
R_{\rho\sigma}\right)F'(\GG)\\
&+4[\nabla_\rho\nabla_\nu F'(\GG)]R_\mu^\rho-4g_{\mu\nu}[\nabla_\rho\nabla_\sigma F'(\GG)]R^{\rho\sigma}-4[\nabla_\rho\nabla_\sigma F'(\GG)]g^{\alpha\rho}g^{\beta\sigma}R_{\mu\alpha\nu\beta}\\
&-2[\nabla_\mu\nabla_\nu F'(\GG)]R+2g_{\mu\nu}[\square F'(\GG)]R-4[\square F'(\GG)]R_{\mu\nu}+4[\nabla_\mu\nabla_\nu F'(\GG)]R_\nu^\rho=\kappa T_{\mu\nu}^{matter},
\end{split}
\end{equation}
where $T_{\mu\nu}^{matter}$ is the energy-momentum tensor.
Taking the flat FLRW  metric, the field equations lead to
\begin{eqnarray}\label{flrw1}
 6H^2+F(\GG)-\GG F'(\GG)+24H^3\dot{\GG}F''(\GG)=2\kappa^2\rho_{matter},
\end{eqnarray}
\begin{equation}\label{flrw2}
\begin{split}
  4\dot{H}&+6H^2+F(\GG)-\GG F'(\GG)+16H\dot{\GG}(\dot{H}+H^2)F''(\GG)\\&+8H^2\ddot{\GG}F''(\GG)+8H^2\dot{\GG}^2F'''(\GG)=
  -2\kappa^2p_{matter},
 \end{split}
\end{equation}
where dots indicate derivation with respect to $\GG$.


 Given a scale factor $a(t)$, we will be able to solve the Friedmann equations in order to see if the model can be realized in the context of Gauss-Bonnet modified gravity.
\subsection[Reconstructing method]{Reconstruction of the solution}
Following \cite{Cognola,Cognolab,Nojiri3}, we want to find a model with $\rho_{matter}=0$ such that the scale factor is a given function, $a(t)$.
Let $t$ be the cosmic time, $P(t)$ and $Q(t)$ arbitrary functions, and consider the action \begin{eqnarray}\label{acciopq}
 S=\frac{1}{2\kappa^2}\int d^4x\sqrt{-g}[R+P(t)\GG+Q(t)].
\end{eqnarray}
Varying with respect to $t$,
\begin{equation}\label{dift}
\frac{dP(t)}{dt}\GG+\frac{dQ(t)}{dt}=0,
\end{equation}
we find $t=t(\GG)$, and evaluating in (\ref{acciopq}) gives $F(\GG)=P(t)\GG+Q(t)$.

Setting $F(\GG)=P(t)\GG+Q(t)$ in the first FLRW equation (\ref{flrw1}), with $\rho_{matter}=0$, and using the energy conservation equation $\dot{\rho}+3H(\rho+p)=0$, we obtain the relation between $P(t)$ and $Q(t)$, as
\begin{eqnarray}\label{eqq}
 Q(t)=-6H^2(t)-24H^3(t)\frac{dP(t)}{dt},
\end{eqnarray}
and replacing $F(\GG)=P(t)\GG+Q(t)$ in this equation yields
\begin{eqnarray}\label{eqasol}
2H^2(t)\frac{d^2P(t)}{dt^2}+2H(t)\left(2\dot{H}(t)-H^2(t)\right)\frac{dP(t)}{dt}+\dot{H}(t)=0.
\end{eqnarray}
\subsection{Stability criteria for the solutions}\label{estabilitat}
Define $H^2(t)=\tilde{g}(N)$, where $N=\ln(a/a_*)$, with $a_*=a(t_*)$ the scale factor at the fiducial time $t_*$, and write $\GG=12\tilde{g}(N)(\tilde{g}'(N)+2\tilde{g}(N))$.
We can rewrite the FLRW equation (\ref{flrw1}) considering $\rho_{matter}=0$ as
\begin{equation}\label{flrwg}
\begin{split}
 6\tilde{g}(N)+F(\GG)-12\tilde{g}(N)(\tilde{g}''(N)+2\tilde{g}(N))F'(\GG)\\+288\tilde{g}^2(N)
 \left((\tilde{g}'(N))^2+\tilde{g}(N)\tilde{g}''(N)+4\tilde{g}(N)\tilde{g}'(N)\right)F''(\GG)=0.
\end{split}
\end{equation}
Let $\tilde{g}$ be solution of this equation; if we perturb it, $\tilde{g}(N)=\tilde{g}_0(N)+\delta\tilde{g}(N)$, and evaluate it in the Friedmann equation (\ref{flrwg}), we obtain
\begin{eqnarray}\label{estab}
\mathcal{J}_1\delta\tilde{g}''(N)+\mathcal{J}_2\delta\tilde{g}'(N)+\mathcal{J}_3\delta\tilde{g}(N)=0,
\end{eqnarray}
where
\begin{equation}\label{j1}
\mathcal{J}_1=288\tilde{g}_0^3(N)F''(\GG_0)\\
\end{equation}
\begin{equation}\label{j2}
\mathcal{J}_2=432\tilde{g}_0^2(N)\left((2\tilde{g}_0(N)+\tilde{g}_0'(N))F''(\GG_0)+
8\tilde{g}_0(N)\left((\tilde{g}_0''(N))^2+\tilde{g}_0(N)(4\tilde{g}_0'(N)+
\tilde{g}_0''(N))\right)\right)
\end{equation}
\begin{equation}\label{j3}
\begin{split}
\mathcal{J}_3=&6\big(1+24\tilde{g}_0(N)\big(\left(-8\tilde{g}_0^2(N)+
3(\tilde{g}'(N))^2+6\tilde{g}_0(N)(3\tilde{g}_0'(N))^2+\tilde{g}''(N)\right)F''(\GG_0)\\
+&24\tilde{g}_0(N)(4\tilde{g}_0(N)+\tilde{g}_0'(N))\left((\tilde{g}_0')^2+
\tilde{g}_0(N)(4\tilde{g}_0'(N)+\tilde{g}_0''(N))\right)F'''(\GG_0)\big)\big),
\end{split}
\end{equation}
and $\GG_0=12\tilde{g}_0(N)(\tilde{g}_0'(N)+2\tilde{g}_0(N))$.

Therefore, the stability conditions read: \ $\frac{\mathcal{J}_2}{\mathcal{J}_1}>0$ and $\frac{\mathcal{J}_3}{\mathcal{J}_1}>0$.

\subsection[Super-bouncing cosmology]{Super-bouncing cosmology}
We here develop a new model in the context of super-bouncing cosmology. The super-bounce scale factor is given by \cite{Oikonomou,Ashtekar,Bojowald}
\begin{eqnarray}\label{asuperbounce}
 a(t)=(t_*-t)^{2/c^2},
\end{eqnarray}
where $t_*$ is the Big Crunch time and $c$ is a parameter, $c>\sqrt{6}$.
This scale factor corresponds to the Hubble parameter
\begin{eqnarray}\label{hsuperbounce}
 H(t)=\frac{2}{c^2(t-t_*)}.
\end{eqnarray}
Replacing $H(t)$ in Eq.~(\ref{eqasol}) and solving it we obtain $P(t)$, then using (\ref{eqq}) we obtain $Q(t)$, and solving (\ref{dift}) with these two particular functions we obtain two solutions for the time, leading to
\begin{eqnarray}
 F_{\pm}(\GG)=\frac{c^4t_*^2\GG\pm16\sqrt{3(2-c^2)\GG}}{8(c^2+2)}\\
\end{eqnarray}
when using $F(t)=P(t)t+Q(t)$. The reconstruction method is explained above, in detail.

We have seen in (\ref{estabilitat}) that the solutions are stable when $\frac{J_2}{J_1}>0$ and $\frac{J_3}{J_1}>0$, where $J_1$, $J_2$ and $J_3$ are defined in Eqs.~(\ref{j1}), (\ref{j2}) and (\ref{j3}).

Observe that in the case where $a(t)$ and $H(t)$ are given by (\ref{asuperbounce}) and (\ref{hsuperbounce}), respectively, we can put
\begin{eqnarray}
H(a)=\frac{2}{c^2}a^{-c^2/2},
\end{eqnarray}
and using $e^{-N}=a_0/a$, we obtain $H(N)=\frac{2}{c^2}a_0^{-c^2/2}e^{-Nc^2/2}$, thus $g(N)=H^2(N)=\frac{4}{c^4}a_0^{-c^2}e^{-c^2N}$.

In this particular case of super-bouncing cosmology, and for the solution $F_+(\GG)$, the stability conditions are
\begin{equation}
\begin{split}
 \frac{J_2}{J_1}&=\frac{3}{2}c^2+3>0,\\
 \frac{J_3}{J_1}&=\frac{7c^{10}-18c^8+a_0^{-2c^2}e^{-2Nc^2}\left(-768c^4+4608c^2-6144\right)}{2c^6}.
\end{split}
\end{equation}

For the solution $F_-(\GG)$, the stability conditions are
\begin{equation}
\begin{split}
\frac{J_2}{J_1}&=\frac{3}{2}c^2+3>0\\
\frac{J_3}{J_1}&=\frac{11c^{10}-18c^8-16c^6+a_0^{-2c^2}e^{-2Nc^2}\left(-768c^4+4608c^2-6144\right)}
{2c^6}.
\end{split}
\end{equation}

We see that $\frac{J_2}{J_1}>0$ in both cases $F_\pm(\GG)$, without any condition for $c$ and for all $t$. In addition, as is clear from Figs.~(\ref{estab1}) and (\ref{estab2}),
if $c>\sqrt{6}$, then $\frac{J_3}{J_1}>0$ when $N\to\infty$, in the models $F_+(\GG)$ and $F_-(\GG)$. As a result, we have successfully constructed a Gauss-Bonnet modified gravity model which realizes the super-bouncing cosmology and which is stable when $t\to\infty$.
\imatge{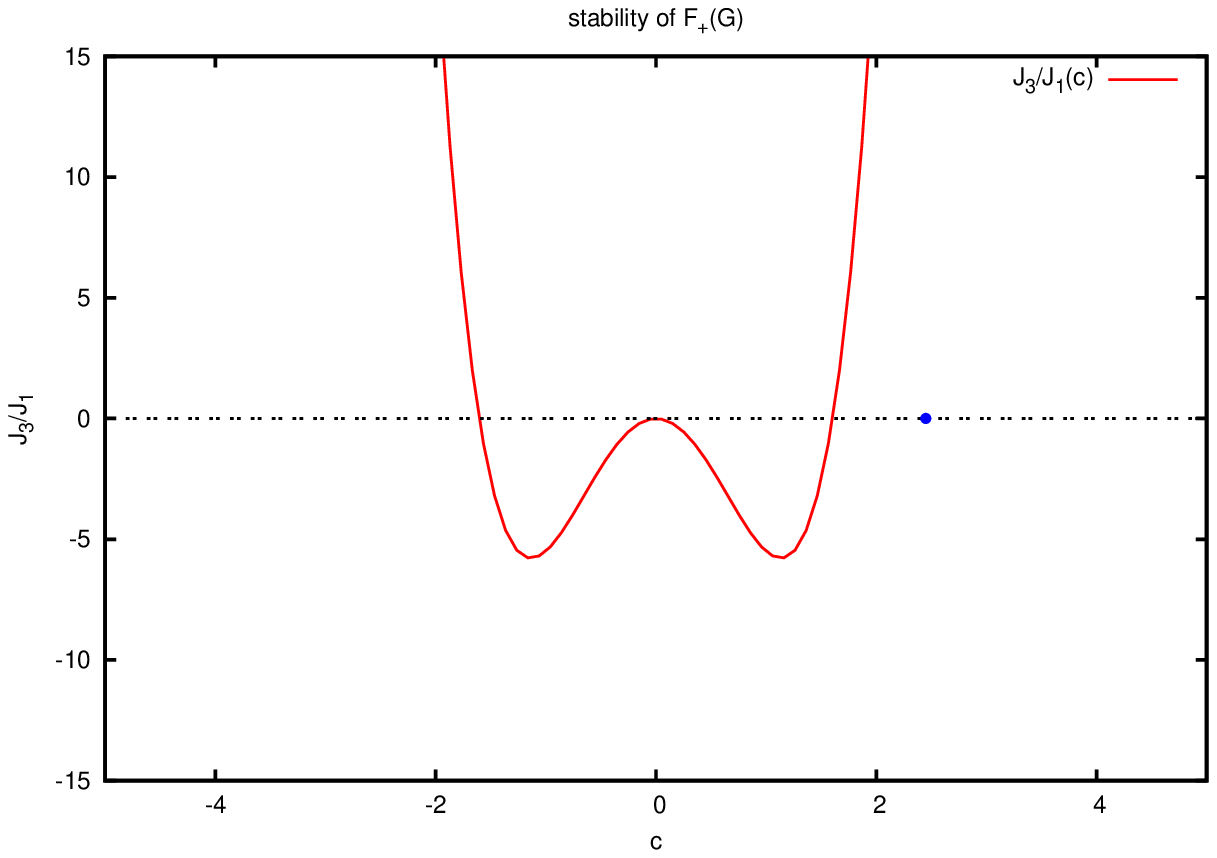}{1.2}{0}{A plot of $\frac{J_3}{J_1}(c)$ for the model $F_+(\GG)$ (red line), where we have taken $a_0=1$ when $N\to\infty$. The blue point corresponds to $c=\sqrt{6}$.}{estab1}
\imatge{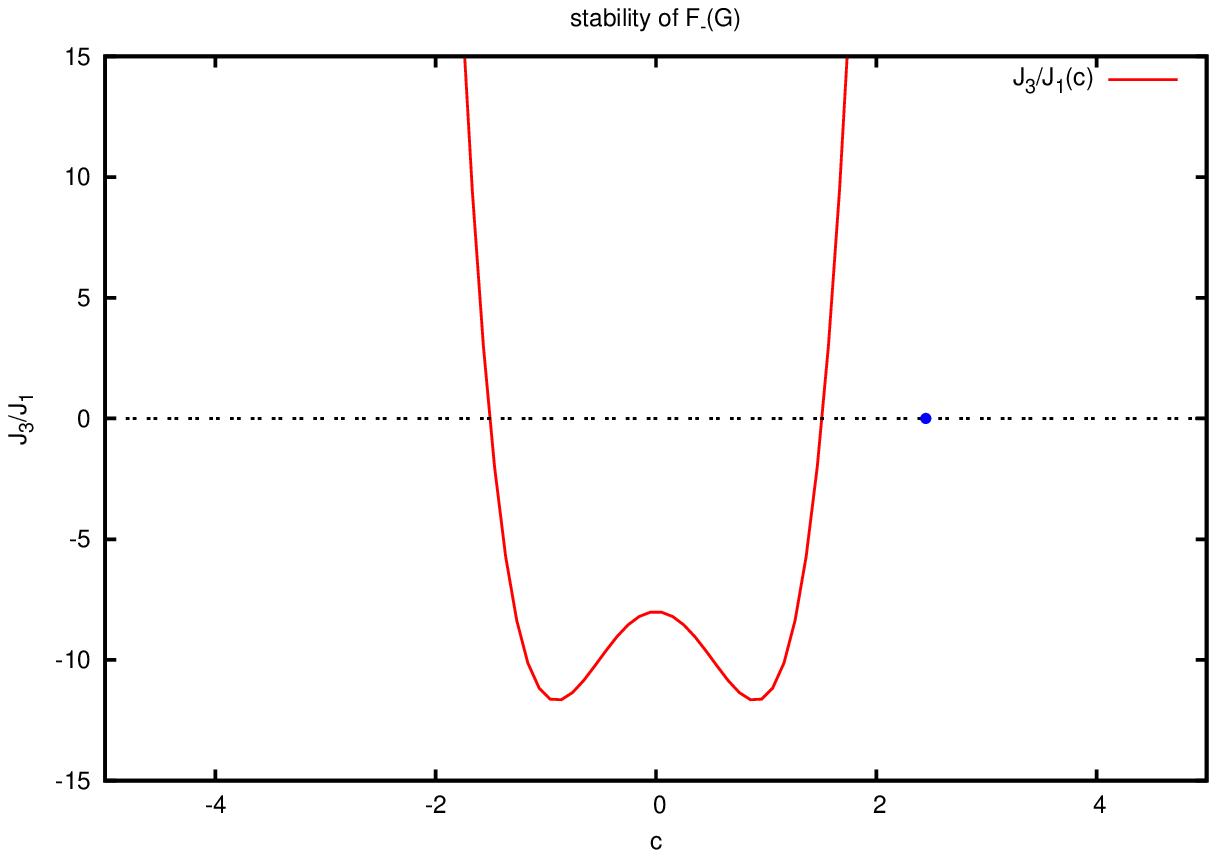}{1.2}{0}{A plot of $\frac{J_3}{J_1}(c)$ for the model $F_-(\GG)$ (red line), where we have taken $a_0=1$ when $N\to\infty$. The blue point corresponds to $c=\sqrt{6}$.}{estab2}

\subsection{Loop Quantum Ekpyrotic cosmology}

As last model, we show here the power of our method in the context of loop quantum ekpyrotic cosmology. In this case, the bouncing scale factor \cite{Oikonomou} reads
\begin{eqnarray}
 a(t)=(a_0t^2+1)^{\rho/2},
\end{eqnarray}
where $0<\rho<1$ is a parameter.
This scale factor corresponds to the Hubble parameter
\begin{eqnarray}
 H(t)=\frac{2a_0\rho t}{a_0t^2+1},
\end{eqnarray}
where $a_0=\frac{8\pi G\rho_c}{3\rho^2}$ and $\rho_c$ is the critical density.
In the limit $t\to\infty$,
\begin{equation}
 a(t)\approx a_0t^\rho,~~~~~~H(t)\approx \frac{\rho}{t}.
\end{equation}
Replacing $H(t)$ in Eq.~(\ref{eqasol}), and solving it we obtain $P(t)$, then using (\ref{eqq}) we obtain $Q(t)$, and solving (\ref{dift}) with these two particular functions we obtain two solutions for the time. Evaluating $F(t)=P(t)t+Q(t)$ at this time, we obtain two models, namely
\begin{eqnarray}
 F_\pm(\GG)=\pm\frac{\sqrt{6\rho\GG(\rho-1)}}{\rho+1}.
 \end{eqnarray}
If $a_0=1$ and when $N\to\infty$, the stability conditions are, in both cases,
  \begin{eqnarray}
\frac{J_2}{J_1}=\frac{3\rho^2-27}{\rho^2+3\rho}
\end{eqnarray}
and
\begin{eqnarray}
\frac{J_3}{J_1}=\begin{cases}
\infty,~~~~&\text{if }\Phi>0,\\
-\infty,~~~~ &\text{if }\Phi<0,
\end{cases}
\end{eqnarray}
where $\Phi=18432\rho^{12}+64512\rho^{11}-82944\rho^9$.
As we can see in Fig.~\ref{estabekpy1}, if $0<\rho<1$, then $\frac{J_2}{J_1}<0$ when $N\to\infty$. As a consequence, we have obtained a new Gauss-Bonnet modified gravity model which realizes the ekpyrotic scenario and which is not stable when $t\to\infty$.
\imatge{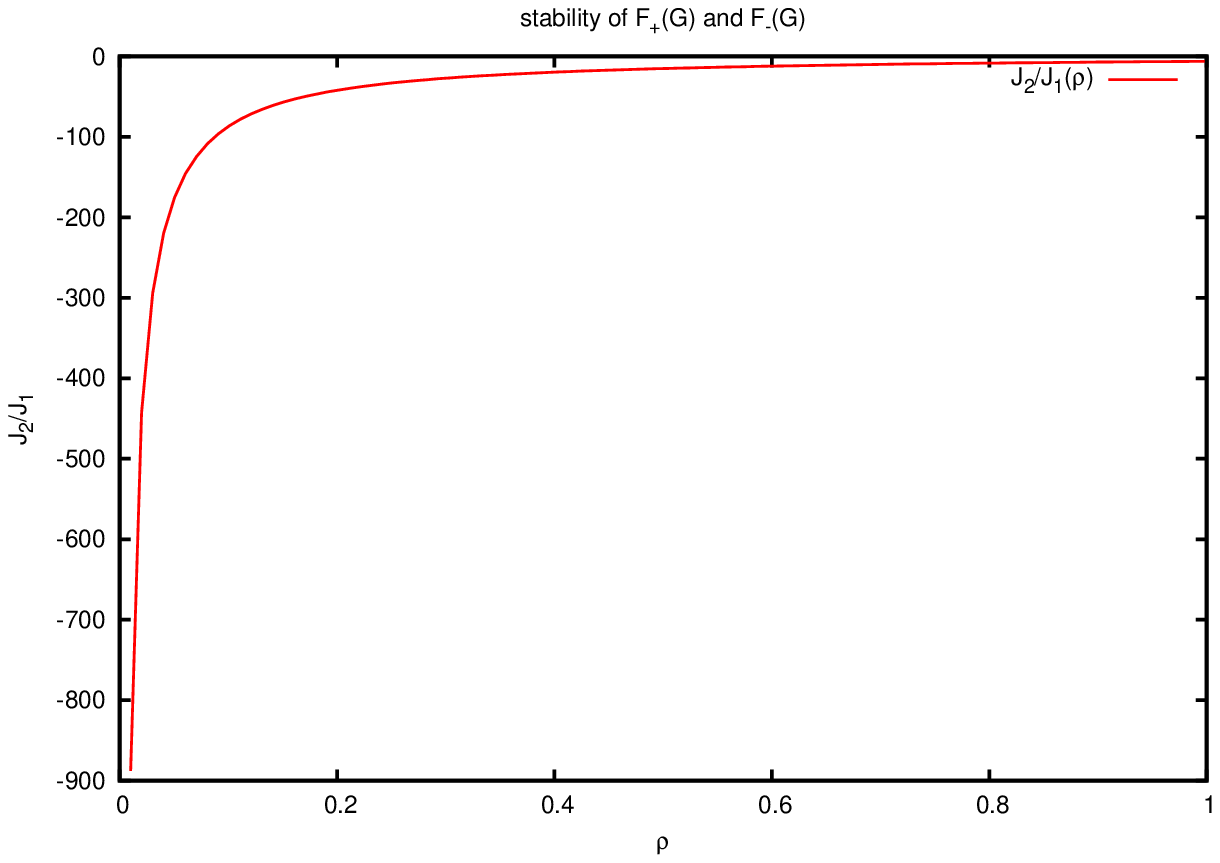}{1.2}{0}{Plot of $\frac{J_2}{J_1}(c)$ for the model $F_\pm(\GG)$, taking $a_0=1$ when $N\to\infty$.}{estabekpy1}
The procedure developed in this paper is quite powerful and can be also applied to get a bounce cosmology with a type IV finite-time singularity \cite{Odintsov3}. Further extensions of the method will be discussed elsewhere.

\section{Conclusions}

The main issue addressed in this paper was how the addition of a Gauss-Bonnet term to some viable models can change the specific properties, and even the physical nature, of their corresponding cosmological solutions. As is known, the addition of a Gauss-Bonnet is very well justified, on  mathematical grounds, moreover, this term is generically obtained from most fundamental theories, as string and M theories. After revisiting several illustrative examples some brand new original dark energy models with quite interesting properties have been constructed which exhibit, in a unified way, the three distinguished possible cosmological phases corresponding to phantom matter, quintessence, and ordinary matter, respectively.

Indeed, in the two first new dark energy models considered in Sect.~2, when $V=0$ and in the case $\gamma=-1$,  the  three  usual cosmological phases are clearly distinguished, which correspond to phantom matter ($w<-1$) when $h_0<0$, to quintessence ($-1<w<-1/3$) when $h_0>1$, and to ordinary matter ($w>7/3$) when $0<h_0<1/5$. In the case $V\neq 0$ it is possible, moreover, to find solutions with constant $w$.
We have also introduced a model in Sect.~2 in which $w$ is a function of time, so that, when the curvature is small, we get $w<-1$ while, when it is large, $w>-1$. In this case, we can either obtain a singularity of the Big Rip kind or a bouncing solution tending to a de Sitter universe with $w=-1$. The fact that such different cosmological behaviors can be unified in this way, as coming from a single cosmological model, need to be properly remarked as a distinguishable success. And also, the important fact that the current acceleration of the universe expansion can be explained by means of a dark energy model which is inspired by, and might be derived from, a truly fundamental theory of physics, as is string theory.

In the second part of the paper, Sect.~3, we have constructed a Gauss-Bonnet modified gravity models with a bouncing behavior in the early stages of the universe evolution, and tested the stability of the corresponding solutions.
A superbounce model has been presented which is stable, as well as a model for the ekpyrotic scenario which is not.

In summary, a number of interesting models have been here studied as alternatives to the standard Big Bang scenario, which could provide a natural and reasonable explanation of the current accelerated expansion of our universe. We have carefully discussed the stability of their solutions and their physical validity. The fact that these well-behaved models may be derived from fundamental physics and are constrained by rigorous mathematical principles (the Gauss-Bonnet invariance) gives a rigorous foundation and an added value to the present and prospective future research along the same lines.
\vspace{5mm}

\noindent{\bf Acknowledgements.}
This investigation has been supported in part by MINECO (Spain), project FIS2013-44881, and by the CPAN Consolider Ingenio Project.

\end{document}